\documentstyle[prd,twocolumn,aps,epsf]{revtex}
\newcommand{\singlefig}[2]{
\begin{center}
\begin{minipage}{#1}
\epsfxsize=#1
\epsffile{#2}
\end{minipage}
\end{center}}
\newcommand{\segmentfig}[3]{
\begin{minipage}{#1}
\epsfxsize=#1
\epsffile{#2}
\begin{center}
{\small \mbox{#3}}
\end{center}
\end{minipage}}
\newcommand{\beq}{\begin{equation}}
\newcommand{\eeq}{\end{equation}}
\newcommand{\beqa}{\begin{eqnarray}}
\newcommand{\eeqa}{\end{eqnarray}}
\newcommand{\bea}{\begin{array}}
\newcommand{\ena}{\end{array}}
\begin{document}
\draft
\title{Black hole solutions in Euler-Heisenberg theory}
\author{Hiroki Yajima\thanks{electronic
mail:yajima@gravity.phys.waseda.ac.jp} and
Takashi Tamaki\thanks{electronic mail:tamaki@gravity.phys.waseda.ac.jp}}
\address{Department of Physics, Waseda University,
Shinjuku, Tokyo 169-8555, Japan}
\date{\today}
\maketitle
\begin{abstract}

We construct static and spherically symmetric 
black hole solutions in the Einstein-Euler-Heisenberg 
(EEH) system which is considered as an effective action of a 
superstring theory. We considered 
electrically charged, magnetically charged and dyon solutions. 
We can solve analytically for the magnetically charged case. 
We find that they have some remarkable properties about causality 
and black hole thermodynamics depending on the coupling constant 
of the EH theory $a$ and $b$, though they have central singularity 
as in the Schwarzschild black hole. We restrict $a>0$ because 
it is natural if we think EH theory as a low energy limit of the 
Born-Infeld (BI) theory. 

(i) For the magnetically charged case, whether or not the 
extreme solution exists depends on the critical parameter $a=a_{crit}$. 
For $a\leq a_{crit}$, there is an extreme solution as in 
the Reissner-Nortstr\"om (RN) solution. 
The main difference from the RN solution is that 
there appear solutions below the horizon radius of the extreme solution 
and they exist till $r_H \to 0$. 
Moreover, for $a> a_{crit}$, there is not an extreme solution. 
For arbitrary $a$, the temperature diverges in the $r_H \to 0$ limit. 

(ii) For the electrically charged case, 
the inner horizon appears under some critical mass $M_0$ 
and the extreme solution always exists. 
The lower limit of the horizon radius decreases 
when the coupling constant $a$ increases. 

(iii) For the dyon case, we expect variety of properties because of 
the term $b(\epsilon_{\mu\nu\rho\sigma}
F^{\mu\nu}F^{\rho\sigma })^{2}$ which is peculiar 
to the EH theory. But their properties are mainly decided by 
the combination of the parameters $a+8b$. We show that 
solutions have similar properties to the magnetically 
charged case in the $r_{H} \to 0$ limit for $a+8b \leq 0$. 
For $a+8b >0$, it depends on the parameters $a,b$.  
\end{abstract}
\pacs{04.70.-s, 04.40.-b, 95.30.Tg. 97.60.Lf.}
%\baselineskip=10mm

%%%%%%%%%%%%%%%%%%%%%%%%%%%%%%%%%%%%%%%%%%%%%%%%%%
\section{Introduction}                            %
%%%%%%%%%%%%%%%%%%%%%%%%%%%%%%%%%%%%%%%%%%%%%%%%%%

Recently, much attention has been payed to BI 
type of actions after 
its recognition as an effective theory of superstring theory\cite{Frad}. 
Moreover, since they describe the action of the Brane, their importance 
has been increasing\cite{Leigh}. In this context, 
there are some studies that 
investigated black hole solutions in the Einstein-BI type  
actions\cite{Oli}. 
Actually a new non-linear
electromagnetism was proposed, which produces a nonsingular
exact black hole solution satisfying the weak energy 
condition\cite{Gar,Eloy}, and have distinct properties
from Bardeen black holes\cite{Bardeen}.
But there remain subjects which should be manifested 
such as thermodynamical properties. 
Here, we concentrate on the EH 
action which first was proposed in 1936\cite{Euler}. 
Though not so much attention has been payed it compared with the BI action, 
EH action well approximate the supersymmetric system of minimally coupled 
spin-$1/2$, -$0$ particles for appropriate parameters\cite{Bern}. 
From the experimental aspects, this is a more accurate 
classical approximation 
of QED than Maxwell's theory when fields have high intensity \cite{Stehle}.  

We investigate the black hole solutions in the EEH system 
from following aspects : (i) The electric-magnetic duality 
(ii) The black hole thermodynamics 
(iii) The causality and stability of the black holes. 
As for the electric-magnetic duality, it is already pointed out 
that though BI action preserves this, EH action breaks it at the 
higher order of the electromagnetic field\cite{Gibbons}. 
Thus black hole solution with electric charge or magnetic charge will have 
clear differences which should be clarified. 
Moreover, the dyon solution may have specific properties which 
can not be seen in the electrically charged or magnetically charged cases. 
The thermodynamical properties of black holes 
are one of the main topics in 
superstring theory after the discoveries of the microscopic origin of the 
black hole entropy\cite{Strominger} and 
the holographic principle\cite{Maldacena}.
It is worth noting that BI type action also plays an 
important role in AdS/CFT 
correspondence\cite{Callan}. 
Causality for the black hole in the EEH system has already 
been investigated by Oliveira.
But this is restricted to the black hole with electric charge, and 
the physical implications such as the stability 
of the black hole are not discussed. 
The stability of the black hole can be  
interpreted from thermodynamical properties established in 
\cite{Kaburaki} and we refer to this by calling it 
a turning point method. 
Using thermodynamical variables, we can easily apply catastrophe theory 
to hairly black holes and this is consistent 
with linear perturbation analysis 
and the turning point method\cite{Torii}. 
We are interested in the thermodynamical properties of black holes in this system 
and the relation between Causality and the stability of black holes. 

This paper is organized as follows.  In Sec. II, we introduce 
basic Ans\"{a}tze and the field  equations in the EEH system. In
Sec. III, we investigate the thermodynamical properties of black holes  
with electric charge or magnetic charge. In Sec. IV, we 
investigate those of dyonic ones.  In Sec. V, we summerize the results 
and comment on future work. Throughout this
paper we use  units $c=\hbar =1$. Notations and definitions 
such as Christoffel symbols  and curvature follow
Misner-Thorne-Wheeler\cite{MTW}.   

%%%%%%%%%%%%%%%%%%%%%%%%%%%%%%%%%%%%%%%%%%%%%%%%%%%%%%%%%%%%
\section{Basic equations}  
%%%%%%%%%%%%%%%%%%%%%%%%%%%%%%%%%%%%%%%%%%%%%%%%%%%%%%%%%%%%

We take the following EEH action 
%%%%%%%%%
\beqa
S = \int d^{4}x\sqrt{-g}\left[\frac{1}{16\pi}
\left(\frac{R}{G}-P+a P^{2}+bQ^{2}
\right) \right],
\eeqa
%%%%%%%%%
where $R$ is the scalar curvature, $P\equiv F^{\mu\nu}F_{\mu\nu}$, 
$Q\equiv \epsilon_{\mu\nu\rho\sigma} F^{\mu\nu}F^{\rho\sigma }$ 
and $\epsilon^{\mu\nu\rho\sigma}$ 
is a completely antisymmetric unit tensor, which yields
%%%%%%%%%
\beqa
\epsilon_{\mu\nu\rho\sigma}\epsilon^{\mu\nu\rho\sigma}=-4!.
\eeqa
%%%%%%%%%
In \cite{Euler}, this corresponds to the weak field approximation and 
the coupling constants are written as 
$a=he^{4}/(360\pi^{2}m^{4})$, $b= 7he^{4}/(1440\pi^{2}m^{4})$, 
where $h$, $e$ and $m$ are the Planck constant, electron charge and 
electron mass, respectively. 
From the present point of view, they should be related to the 
inverse string tension $\alpha '$ which restrict $a>0$ 
because of its correspondence to the BI action in the low energy limit. 
As has been pointed out in \cite{Frad}, to construct gravitational 
counterpart of the BI action involves much difficulty. Here, we adopt the 
Einstein-Hilbert action as a first approximation.  
We can derive Einstein equations as 
%%%%%%%%%
\beqa
G_{\mu\nu}&=& \frac{1}{2}g_{\mu\nu}(-P+aP^{2}+bQ^{2}) 
+2F_{\mu\lambda}F_{\nu}^{\lambda}\nonumber \\
&&-4aP (F_{\mu\lambda}F_{\nu}^{\lambda})-
8bQ (\epsilon_{\mu\zeta\eta\vartheta}F^{\zeta\eta}F_{\nu}^{\vartheta}).
\eeqa
%%%%%%%%%
We consider the metric of static and spherically symmetric, 
%%%%%%%%%
\beqa
ds^{2}= -f(r) e^{-2\delta(r)} dt^{2}+
f(r)^{-1}dr^{2}+r^{2}d\Omega^{2},
\eeqa
%%%%%%%%%
where $f(r)\equiv 1-2Gm(r)/r$. 
We introduce the gauge potential $A_{\mu}$, as 
%%%%%%%%%
\beqa
A_{\mu}=(A(r), 0, 0, Q_{m}\cos\theta).
\eeqa
%%%%%%%%%
Then, the Einstein equations are 
%%%%%%%%%
\beqa
-\frac{2 G m'}{r^2}= - F_{e}-F_{m}-F_{dy}
\label{eqn:g00},
\eeqa
%%%%%%%%%
\beqa
-\left(\frac{2 G m'}{r^2}+\frac{2}{r}\delta ' 
f\right)= -F_{e}-F_{m}-F_{dy}
\label{eqn:g11},
\eeqa
%%%%%%%%%
where $'$ represents $d/dr$. 
We used the abbreviations as 
%%%%%%%%%
\beqa
F_{e}&\equiv & e^{2\delta}(A')^{2}+6a e^{4\delta }(A')^{4} , \\
F_{m}&\equiv & \frac{Q_{m}^{2}}{r^{4}}-2a\frac{Q_{m}^{4}}{r^{8}} ,  \\
F_{dy}&\equiv & (96b-4a)e^{2\delta}(A')^{2}\frac{Q_{m}^{2}}{r^{4}} .
\eeqa
%%%%%%%%%
Subtracting Eq. (\ref{eqn:g00}) from Eq. (\ref{eqn:g11}) 
yields $d\delta/dr= 0$. 
We require asymptotically flatness for the solution. 
%%%%%%%%%
\beqa
A(r) \rightarrow -\frac{Q_{e}}{r}, \ \ \delta(r) \rightarrow 0, \ \ 
Gm(r) \rightarrow constant,   
\eeqa
%%%%%%%%%
as $r\rightarrow \infty$. 
Thus we obtain $\delta =0$.
%%%%%%%%%
So we have only one independent Einstein equation as
%%%%%%%%%
\beqa
Gm'=\frac{r^2}{2}\left( F_{e}+F_{m}+F_{dy}\right).
\label{eqn:mass'}
\eeqa
%%%%%%%%%
The field equation is
%%%%%%%%%
\beqa
4 a r^{2} (A')^{3}+A'z(r) = Q_{e},
\label{eqn:field0}
\eeqa
%%%%%%%%%
where
%%%%%%%%%
\beqa
z(r) \equiv r^{2}-4 (a+8b) \frac{Q_{m}^{2}}{r^{2}}.
\eeqa
%%%%%%%%%
This is third order algebraic equation for $A'$ except for $Q_e =0$. 
For regularity at the horizon $r_H$, we require 
%%%%%%%%%
\beqa
Gm(r_{H})=\frac{1}{2}r_{H},\ \ A(r_H)<\infty . 
\eeqa 
%%%%%%%%%

%%%%%%%%%%%%%%%%%%%%%%%%%%%%%%%%%%%%%%%%%%%%%%%%%%%%%%%%%%%%
\section{Black hole solutions with electric or magnetic charge}    %
%%%%%%%%%%%%%%%%%%%%%%%%%%%%%%%%%%%%%%%%%%%%%%%%%%

In this section, we show the properties of black hole solutions with 
magnetic or electric charge. First, we point out that the zeroth and 
the first law of black hole thermodynamics 
can be applicable even for non-linear matter terms which violate 
dominant energy condition though Smarr's formula can not\cite{Heusler}. 

%%%%%%%%%%%%%%%%%
\subsection{Magnetically charged case}
%%%%%%%%%%%%%%%%%

In the case $Q_{e} \equiv 0$, we can solve equations analytically. 
In this case, there remains only $F_{m}$ part in Eq. (\ref{eqn:mass'}). 
Note that $Gm'$ can be negative which makes an intrinsic difference 
from the RN solution. We can integrate Eq. (\ref{eqn:mass'}),
%%%%%%%%%%%%
\beqa
Gm=GM-\frac{Q_{m}^{2}}{2 r}+a \frac{Q_{m}^{4}}{5 r^{5}}, 
\eeqa
%%%%%%%%%%%%
where $M$ is the gravitational mass of the black hole. 
Thus, the horizon radius $r_{H}$ must satisfy 
%%%%%%%%%%%%
\beqa
h(r_{H}) \equiv r_{H}^{6}-2 G M r_{H}^{5}+Q_{m}^{2} r_{H}^{4}-
\frac{2}{5} a Q_{m}^{4} = 0. 
\label{qmclass}
\eeqa
%%%%%%%%%%%%
Since $h(0)<0$ and $h(\infty)\to \infty$, 
the solution which satisfies $h(r_{H})=0$ for $r_H >0$ always exists. 

From (\ref{qmclass}), 
%%%%%%%%%%%%
\beqa
\frac{dh}{dr_{H}}=2r_{H}^{3}(3r_{H}^{2}-5GMr_{H}+2Q_{m}^{2}). 
\label{qmclass}
\eeqa
%%%%%%%%%%%%
So we can classify the number of the horizon as follows. 
For $L\equiv (5GM)^{2}-24 Q_{m}^{2}> 0$, if  $h([5GM+\sqrt{L}]/6) < 0$ 
and $h([5GM-\sqrt{L}]/6) > 0$, 
there are three positive solutions, which means that 
there are one outer horizon and two inner horizons. 
If $h([5GM+\sqrt{L}]/6) = 0$ or 
$h([5GM-\sqrt{L}]/6) = 0$, 
there are two horizons. In other cases, there is only one horizon. 

%%%%%%%%%%%%%%%%%%%%
\begin{figure}
\begin{center}
\singlefig{9cm}{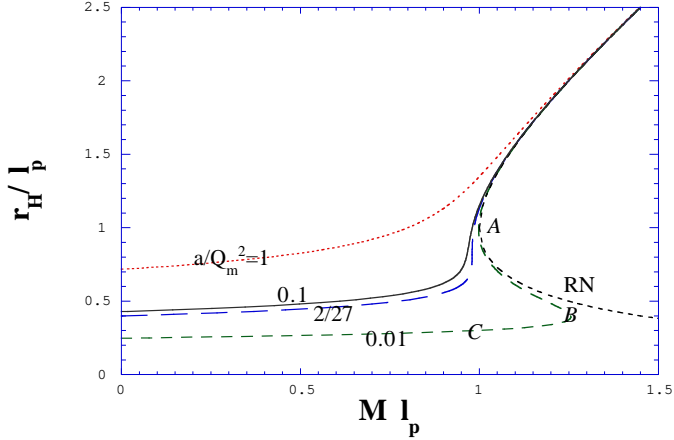}
\caption{$M$-$r_{H}$ relation for $Q_{m}/l_{p}=1$ and $a/Q_{m}^{2}=0$, $0.01$, 
$2/27$, $0.1$, $1$. The points $A$ correspond to the extreme solutions. 
The lines between $A$ to $B$, $B$ to $C$ correspond to outer inner 
horizon and inner inner horizon, respectively. 
We can see that $a_{crit} =2Q_m^2/27$ divides the properties 
qualitatively. 
\label{M-rHqm} }
\end{center}
\end{figure}
%%%%%%%%%%%%%%%%%%%%
\begin{figure}
\begin{center}
\singlefig{9cm}{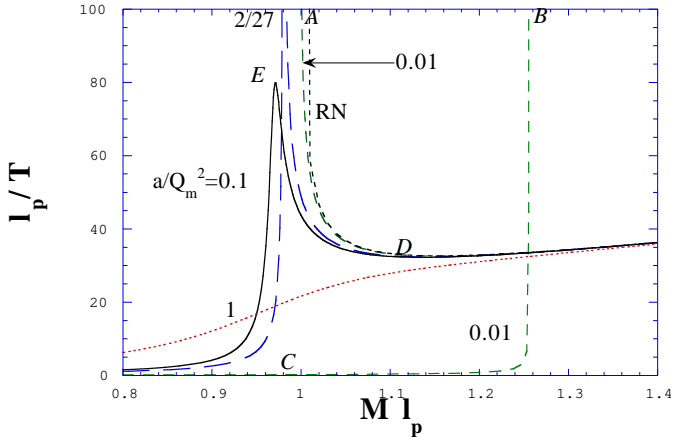}
\caption{$M$-$1/T$ relation for the same parameters in Fig. 1. 
For $a\leq a_{crit}$, there is an extreme solution where 
the temperature is zero. For arbitrary $a$, the 
temperature diverges in the $r_H \to 0$ limit. 
\label{M-1Tqm} }
\end{center}
\end{figure}
%%%%%%%%%%%%%%%%%%%%

We also evaluate that what condition would be required to exist 
an extreme solution. $Gm' =1/2$ leads to 
%%%%%%%%%%%%
\beqa
K(r)\equiv r^{6}-Q_{m}^{2}r^{4}+2aQ_{m}^{4}=0 .
\eeqa
%%%%%%%%%%%%
Thus 
%%%%%%%%%%%%
\beqa
K' = 2r^{3}(3r^{2}-2Q_{m}^{2}) .
\eeqa
%%%%%%%%%%%%
So $r=\sqrt{2/3}Q_{m}$ is a local minimum of $K$ for $r>0$. 
$K(\sqrt{2/3}Q_{m})=0$ leads to $a=(2/27)Q_{m}^{2}\equiv a_{crit}$, 
which means that there is not an extreme solution for $a>a_{crit}$. 

We first show the relation between the gravitational mass $M$ and 
the horizon $r_H$ for $Q_{m}/l_{p}=1$ and $a/Q_{m}^{2}=0$, $0.01$, 
$2/27$, $0.1$, $1$ (Fig. 1). $l_{p}$ is the Planck length. 
For $a/Q_{m}^{2}=0$, $0.01$, $2/27$, there is an extreme solution 
(the point $A$). The lines between $A$ to $B$, $B$ to $C$ 
correspond to the outer 
inner horizon and the inner inner horizon, respectively. 
Note that below the point $C$, there are black hole solutions again. 
For $a/Q_{m}^{2}=0.1$, $1$, there is not an extreme solution 
as we showed. In all cases, $M \to -\infty$ for $r_H \to 0$. 
We also show the gravitational mass $M$ and the 
inverse temperature $1/T$ relation in Fig. 2. 
Above the mass corresponding to the point $A$, 
it is similar to the RN's qualitatively. But below the 
mass corresponding 
to the point $C$ is quite different from RN's. 
The temperature is finite but nonzero at the point $C$. 
The curve from $C$ to $B$ means that if we regard an inner 
horizon as an event horizon, the `temperature' goes to 
zero when $r_H$ approaches the point $B$. 
If we apply the turning point method in this case, the line 
below the point $C$ would be unstable. So we can regard 
this as unphysical. 
But for $a/Q_{m}^{2}=0.1$, $1$, this method suggest that there is no 
stability change if we think in the isolated system though 
thermodynamical properties are different in these two cases. 
The specific heat of the black hole 
never changes for $a/Q_{m}^{2}=1$, while 
it changes twice at the points $D$ and $E$ 
for $a/Q_{m}^{2}=0.1$. 
In all cases, the temperature diverges for $r_H \to 0$. 
It is reasonable that higher order curvature terms would 
change the results in this region. 
But even if we believe that this system describes black hole 
solutions correctly, it is difficult to observe the negative 
mass black holes since it will evaporate very quickly.

%%%%%%%%%%%%%%%%%%%%%%%%%%%%%%%%%%%
\subsection{Electrically charged case}
%%%%%%%%%%%%%%%%%%%%%%%%%%%%%%%%%%%

In this case, from Eq. (\ref{eqn:field0}), 
%%%%%%%%%%%%%
\beqa
4 a r^{2} (A')^{3}+A' r^{2} = Q_{e},
\label{eqn:field1}
\eeqa
%%%%%%%%%%%%%
which has only one real solution and two imaginary solutions.
The real solution is
%%%%%%%%%%%%%
\beqa
A'(r) = \frac{-2\cdot 3^{1/3} r^{2} +2^{1/3}B^{2/3}}
{6^{2/3} \sqrt{x}B^{1/3}}
\label{eqn:field01},
\eeqa
%%%%%%%%%%%%%
We used an abbreviation as
%%%%%%%%%%%%%
\beqa
B \equiv 9 \sqrt{x} Q_e + \sqrt{12 r^{6}+81 x Q_e^2}, \\
x(r)\equiv 4ar^{2} .
\eeqa
%%%%%%%%%%%%%
There remains only $F_{e}$ part in Eq. (\ref{eqn:mass'}) 
which shows $Gm' \geq 0$. This is one of the main difference 
from the magnetically 
charged case. We show that an extreme solution always exists. 
If we take $Gm'(r_H)=1/2$, $A'(r_H)$ is evaluated from 
Eq. (\ref{eqn:mass'}) as
%%%%%%%%%%%%% 
\beqa
A'(r_H) = \frac{(y-1)^{1/2}}{2 (3 a)^{1/2}}\label{eqn:eb3}. 
\eeqa
%%%%%%%%%%%%%
We introduced a dimensionless valiable $y$ as 
%%%%%%%%%%%%% 
\beqa
y = \sqrt{1+\frac{24 a}{r_H^2}} .
\eeqa
%%%%%%%%%%%%%
Substituting Eq. (\ref{eqn:eb3}) into Eq. (\ref{eqn:field01}) derives 
%%%%%%%%%%%%%
\beqa
g(y) \equiv (y-1)^{3/2} + 3 (y-1)^{1/2} - 
\frac{Q_e (y^2-1)}{4 (3 a)^{1/2}}= 0 \ .
\label{eqn:eb4}
\eeqa 
%%%%%%%%%%%%%
Thus, 
%%%%%%%%%%%%%
\beqa
g'(y) = \frac{3}{2}\frac{y}{(y-1)^{1/2}} - \frac{Q_e y}{2 (3 a)^{1/2}}.
\eeqa
%%%%%%%%%%%%%

Because of $y>1$, the solution of $g'(y)=0$ is $y=y_0=1+27a/Q_e^2 (>1)$. 
We can see $g'(y)>0$ for $1<y<y_0$ and $g'(y)<0$ for $y_0<y$. 
So if we notice that $g(1)=0$ and $g\to -\infty \ (y \to \infty)$, 
we find there is only one positive solution. 
So we can conclude that there is one extremal black hole for $a>0$. 

Electrically charged case has already been investigated previously
\cite{compare}. The solution can be expressed using 
the hypergeometric function. But we need numerical 
calculation to investigate their detailed properties, 
particularly their thermodynamical properties. 
The inner horizon only appears for black hole solution 
$M<M_{0}$ as he showed. $M_{0}$ is 
%%%%%%%%%%%%%%%%%
\beqa
M_{0}=\frac{\Gamma (1/4)}{2\Gamma (3/2)}
\frac{Q_{e}^{3/2}}{(2a)^{1/4}}.
\eeqa
%%%%%%%%%%%%%%%%%
We first show the field distributions of the solutions 
((a)$r$-$m$ (b)$r$-$A'$) 
for $r_{H}/l_{p} =1$, $Q_{e}/l_{p}=1$ and $a/Q_{e}^{2}=0$, 
$0.1$, $1$, $10$ in Fig. 3.
Because its difference from the Maxwell field is particularly large near 
the event horizon, this makes a nontrivial change for 
a small black hole. 
We can evaluate that for $a^{1/2}Q_e \gg r^{2}$, 
%%%%%%%%%%%%%%%%%
\beqa
A'\sim \frac{Q_{e}^{1/3}}{(2r)^{2/3}a^{1/3}}.
\label{asymqe}
\eeqa
%%%%%%%%%%%%%%%%%

%%%%%%%%%%%%%%%%%%%%
\begin{figure}[htbp]
\segmentfig{9cm}{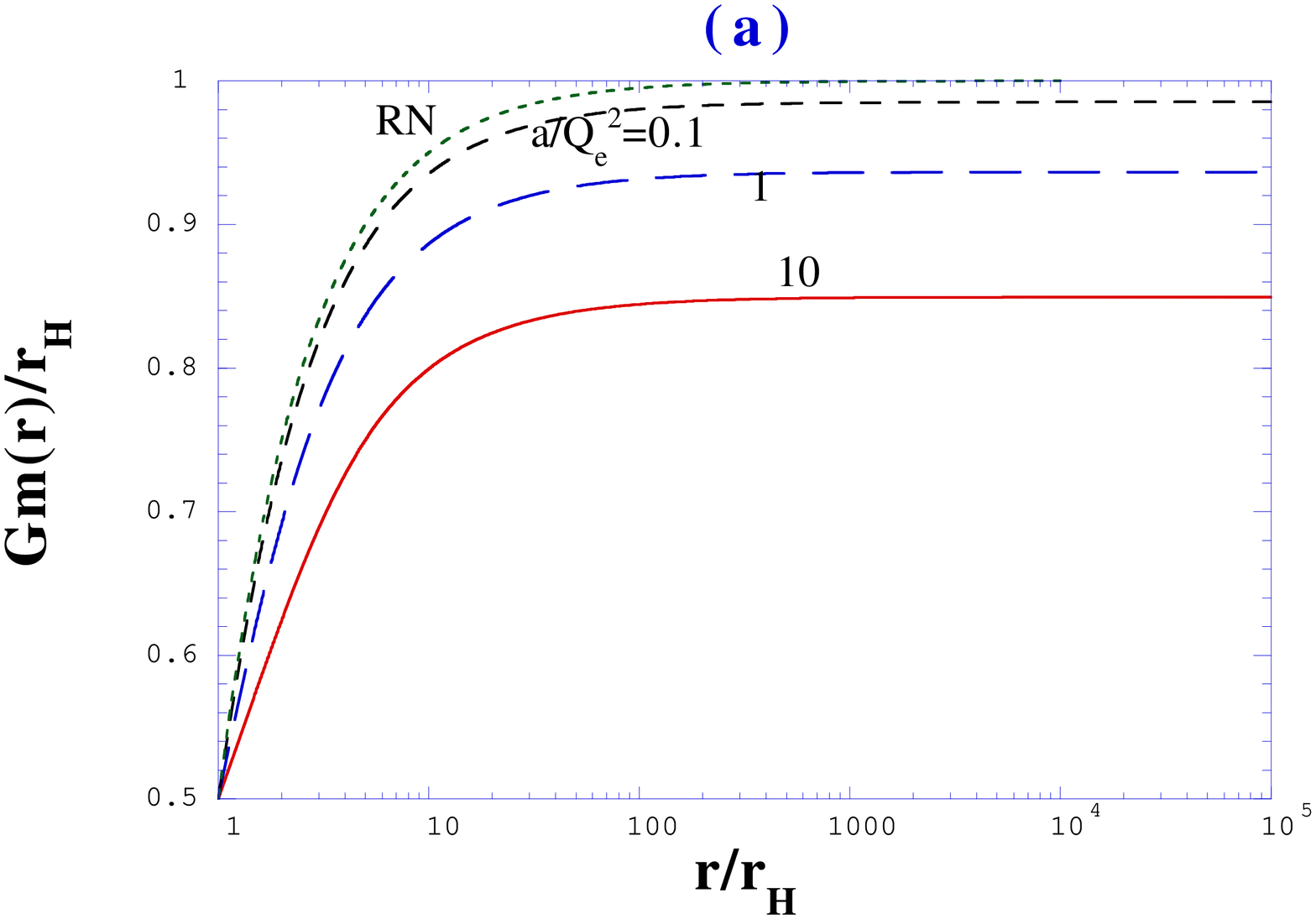}{}  \\
\segmentfig{9cm}{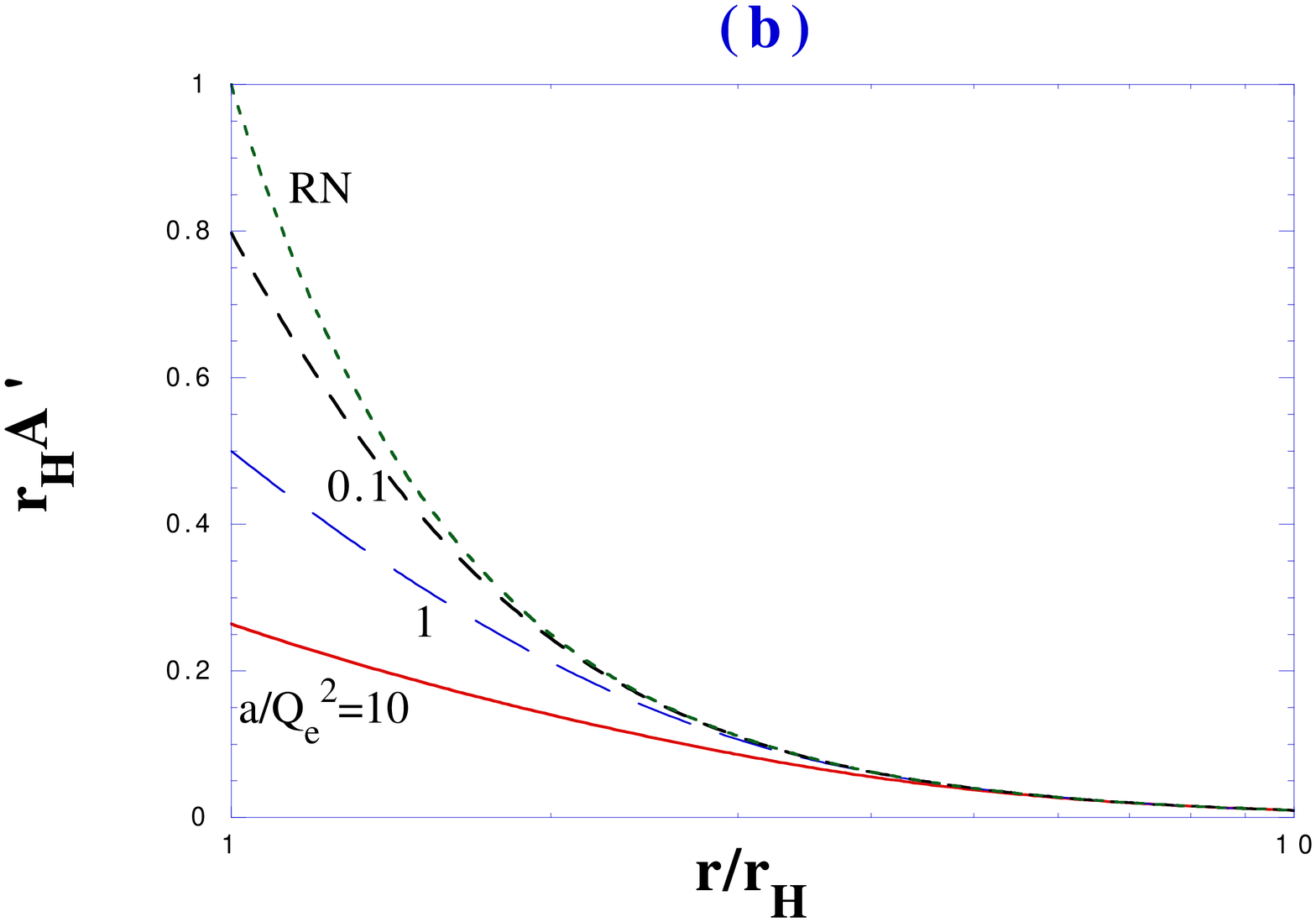}{}
\caption{Field distributions of black holes with electric charge 
for $r_{h}/l_{p}=1$, $Q_{e}/l_{p}=1$ and $a/Q_{e}^{2}=0$, $0.1$, 
$1$, $10$ ((a) $r$-$m$ (b) $r$-$A'$). 
Because of the difference from Maxwell field at small scale, 
the resulting solution deviates from the RN black hole 
near the horizon. $A'$ monotonically decreases as $r \to \infty$ 
as is easily shown. 
\label{field-qe}
}
\end{figure}
%%%%%%%%%%%%%%%%%%%%%

Differentiating Eq. (\ref{eqn:field01}) shows $A''<0$, so 
$A'$ monotonically decreases as $r \to \infty$.  
We investigate the $M$-$r_H$ and 
$M$-$1/T$ relations for electrically charged black holes 
in Fig. 4 and 5, respectively. We take $a/Q_{e}^{2}=0$, 
$0.1$, $1$, $10$. 
The point $A$ corresponds to the extreme solution and 
the curve $A$ to $B$ shows an inner horizon. 
For finite $a$, there exists an 
extreme solution as we noted above and this approaches to $r_H \to 0$ 
for $a \to \infty$. Thermodynamical properties are similar 
to the case for the RN solution. 
The point $D$ corresponds to the point where 
the specific heat changes and this is not equivalent to the point $B$. 
So there is no relation between the point 
which is relevant to the causality change and the point 
at which the specific heat changes. 

%%%%%%%%%%%%%%%%%%%%%%%%
\begin{figure}
\begin{center}
\singlefig{9cm}{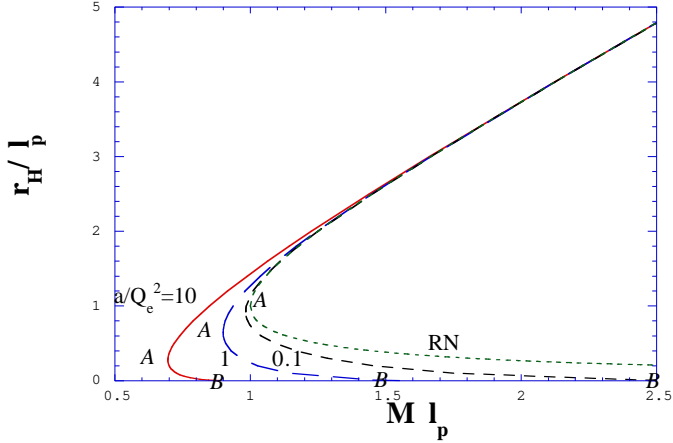}
\caption{$M$-$r_H$ relation for $a/Q_{e}^{2}=0$, $0.1$, $1$, $10$. 
The causality changes at the point $B$ below which 
the inner horizon appears. 
Though the lower limit of the horizon 
decreases as we take $a$ large, the extreme solution always exists. 
\label{M-r_Hele} }
\end{center}
\end{figure}
%%%%%%%%%%%%%%%%%%%%
\begin{figure}
\begin{center}
\singlefig{9cm}{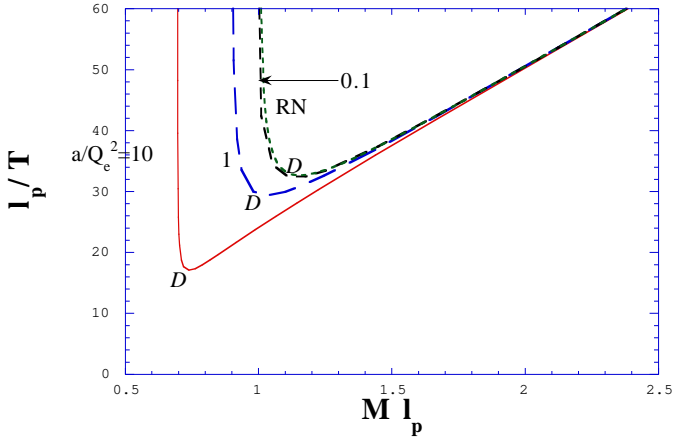}
\caption{$M$-$1/T$ relation for the same parameters in Fig. 4. 
The extreme solution always exists 
where the temperature becomes $0$. So the $M$-$1/T$ 
relation is similar to the one for RN black hole. 
Note that the point where 
the sign of the specific heat changes does not necessarily 
correspond to the point $B$ which suggests that the causality change 
will not be irrelevant to the stability change. 
\label{M-1Tele} }
\end{center}
\end{figure}
%%%%%%%%%%%%%%%%%%%%
 
%%%%%%%%%%%%%%%%%%%%%%%%%%%%%%%%%%%%%%%%%%%%%%%%%%
\section{Dyon black hole}    %
%%%%%%%%%%%%%%%%%%%%%%%%%%%%%%%%%%%%%%%%%%%%%%%%%%

As we showed above, the properties of black holes have 
very different aspects, depending on 
whether it has electric charge or magnetic charge. 
In this section, 
one of the main purposes is to survey how thermodynamical properties 
change when we change the $Q_{m}/Q_{e}$ ratio 
or the coupling constants $a$, $b$. 
In this case, from Eq. (\ref{eqn:field0}), the three solutions 
are expressed as 
\beqa
A'(r) = \frac{-2\cdot 3^{1/3} z +2^{1/3}B^{2/3}}
{6^{2/3} \sqrt{x}B^{1/3}}
\label{eqn:field001},
\eeqa
\beqa
A'(r) = \frac{(1 \pm i \sqrt{3}) x}
{2^{2/3} 3^{1/3} \sqrt{x}B^{1/3}}-
\frac{(1 \mp i \sqrt{3})B^{1/3}}
{2^{4/3} 3^{2/3} \sqrt{x}}
\label{eqn:field02}.
\eeqa
In this case, 
%%%%%%%%%%
\beqa
B = 9 \sqrt{x} Q_e + \sqrt{12 z^{3}+81 x Q_e^2}.
\eeqa
%%%%%%%%%%

%%%%%%%

Note that for $12z^3+81xQ_e^2\geq 0$, the only real solution is 
(\ref{eqn:field001}). For $a+8b>0$, because $z$ can be negative 
for a small $r$, $12z^3+81xQ_e^2< 0$ is possible only 
near the horizon. 
But even in that case, there is only one positive solution 
(\ref{eqn:field001}). 
We should take a positive solution because $z$ eventually 
becomes positive for large values of $r$. 
So we take (\ref{eqn:field001}) in any case.

We can classify solutions in the $r_{H} \to 0$ limit three types as follows. 

(I) If $a+8b=0$, $A'(r)$ approach (\ref{asymqe}) for $a^{1/2}Q_e \gg r^2$, 
which is the same as in the electrically charged case. On the contrary, 
Eq. (\ref{eqn:mass'}) approach 
%%%%%%%%%%
\beqa
Gm' \sim -a\frac{Q_{m}^4}{r^6},
\label{likeqm}
\eeqa
%%%%%%%%%%
which has same form as in the magnetically charged case. 
So the characteristic feature of small $r_H$ is like that of 
the magnetically charged case. 

For $a+8b \neq 0$, we can see its nature if 
we rewrite (\ref{eqn:field001}) as 
%%%%%%%%%%
\beqa
A' = \frac{A_{1}^{3}-A_{2}^{3}}
{6^{2/3}\sqrt{x}B^{1/3}(A_1^2+A_1 A_2+A_2^2)},
\eeqa
%%%%%%%%%
where 
%%%%%%%%%
\beqa
A_{1}\equiv 2^{1/3}B^{2/3}, \ A_{2}\equiv 2\cdot 3^{1/3}z . 
\eeqa
%%%%%%%%%

(II) For $a+8b< 0$, we can evaluate 
%%%%%%%%%
\beqa
A_1^3-A_2^3 = 36Q_e\sqrt{x}B ,
\eeqa
%%%%%%%%%
which shows $A' \propto r^{2}$ in the $r \to 0$ limit. 

So we can conclude that if $(a+8b) \leq 0$, Eq. (\ref{eqn:mass'}) 
has same asymptotically form (\ref{likeqm}) in the $r_{H} \to 0$ 
limit as in the magnetically charged case. 

(III) For $a+8b > 0$, there exists $r = r_{0}$ below which 
$12z^3+81Q_e^2 x<0$ is satisfied. 
For a while we consider $r<r_{0}$ case. 
Then we can evaluate 
%%%%%%%%%%
\beqa
A_1^3-A_2^3 = -24z^3, 
\eeqa
%%%%%%%%%%
which shows $A' \propto r^{-2}$ in the $r \to 0$ limit. 
So we can not conclude 
whether or not solutions in the $r_{H}$ limit exists. 
It depends on $a$, $b$ as we see below. 

Next, we show the field distributions in Fig. 6 (a) $r$-$m$ (b) $r$-$A'$ 
for $a/Q_{e}^{2}=1$, $b/Q_{e}^{2}=-1$, (i.e., $a+8b<0$), 
$r_{H}/l_{p}=1$, $Q_{e}/l_{p}=1$ and 
$Q_{m}/l_{p}=10^{-4}$, $1$. Monotonically decrease of 
$A'$ is broken and $m'<0$ region is specific for $Q_{m}/l_{p}=1$ 
contrary to the case for $Q_{m}/l_{p}=10^{-4}$. 
But they are universal in the $r_{H}\to 0$ limit unless 
$Q_{m}\neq 0$ as is shown above. 

%%%%%%%%%%%%%%%%%%%%%%%
\begin{figure}[htbp]
\segmentfig{9cm}{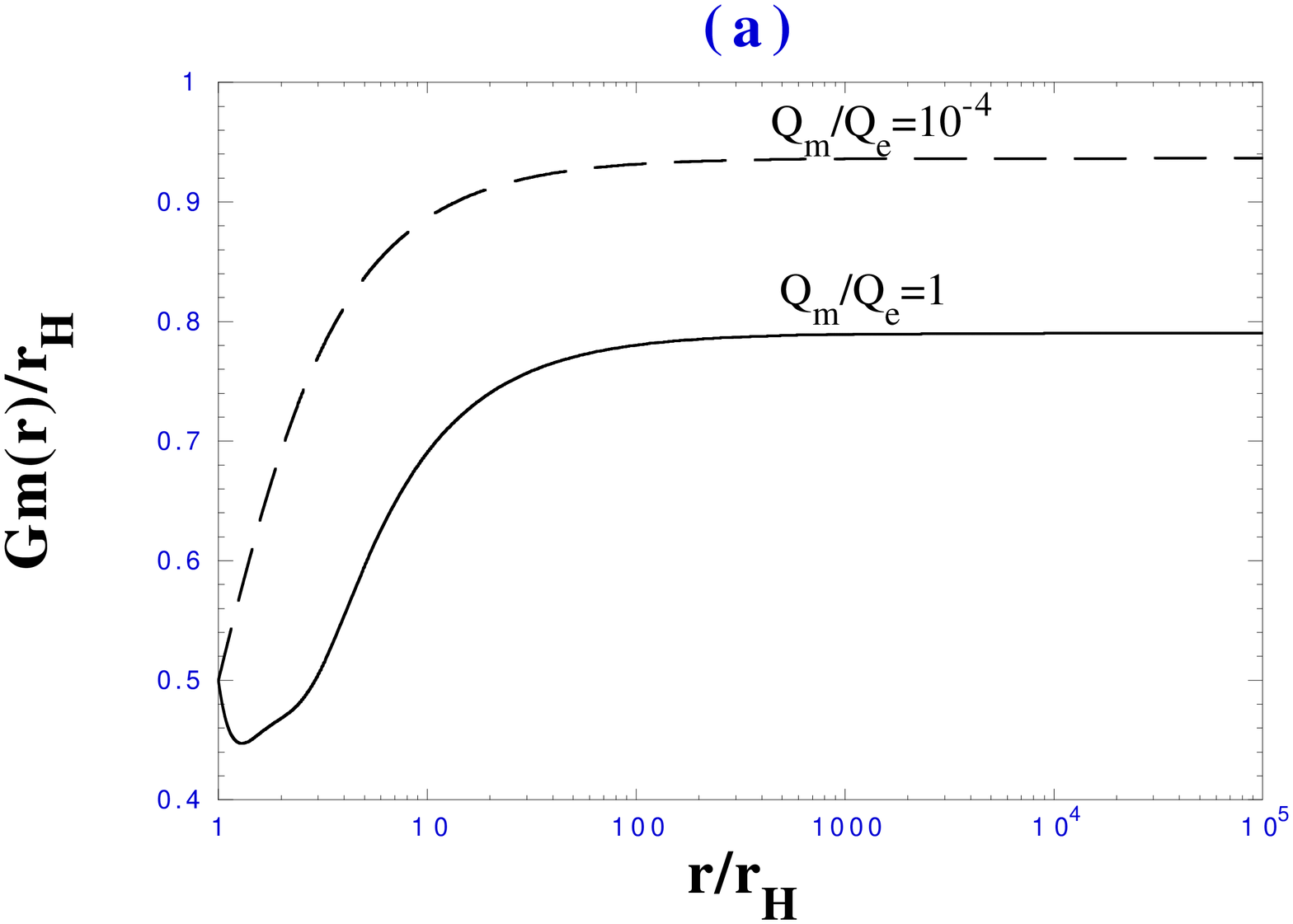}{}  \\
\segmentfig{9cm}{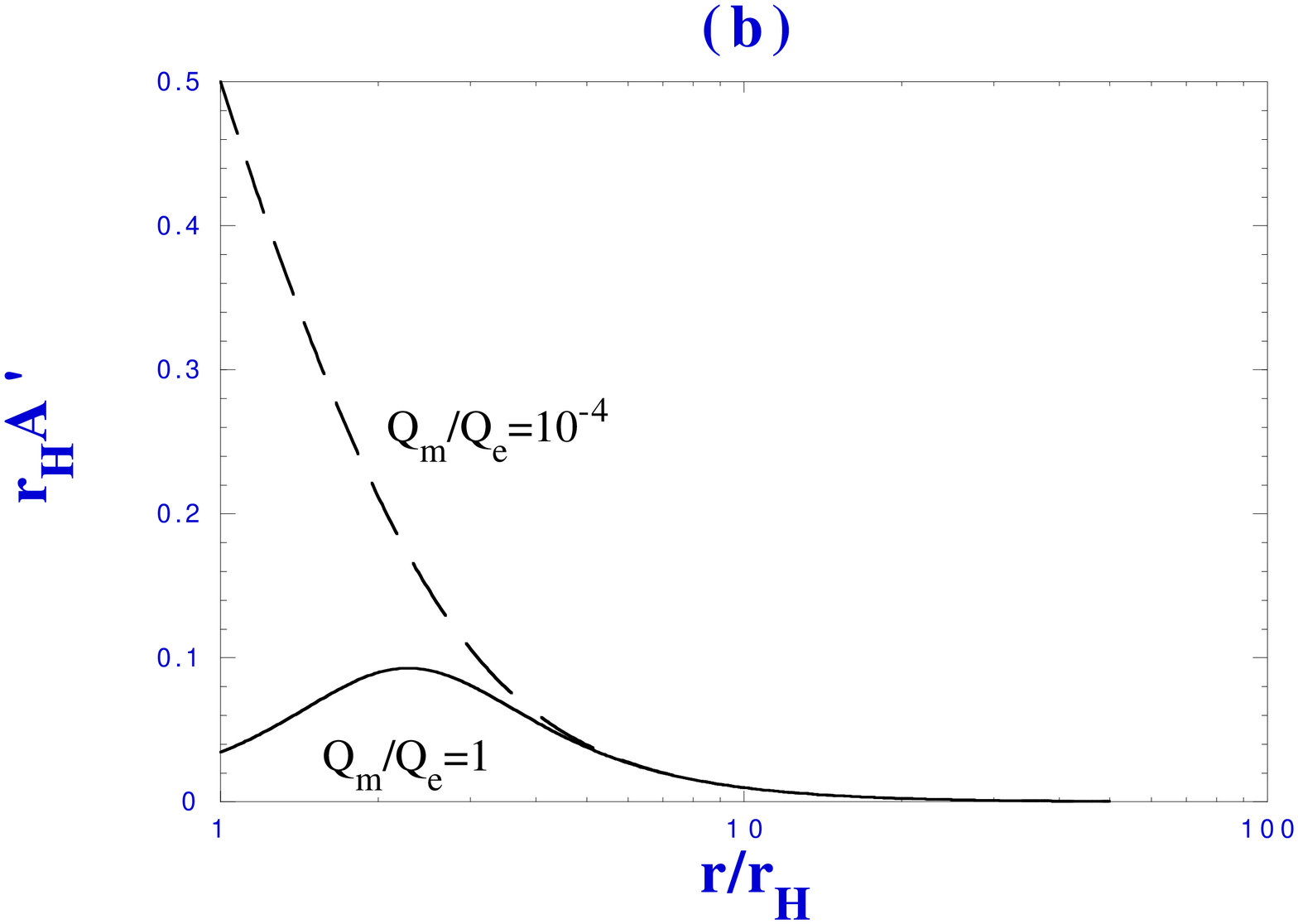}{}
\caption{Field distributions of dyon black holes 
for $r_{H}/l_{p}=1$, $a/Q_{e}^{2}=1$, $b/Q_{e}^{2}=-1$, 
$Q_{e}/l_{p}=1$ and $Q_{m}/l_{p}=10^{-4},1$ 
((a) $r$-$m$ (b) $r$-$A'$). As we can see for $Q_{m}/l_{p}=1$ 
monotonically decreasing of $A'$ is broken and $m'<0$ region 
appears which can be seen in the $r_{H}\to 0$ limit unless 
$Q_{m}\neq 0$. 
\label{field-dyon}
}
\end{figure}
%%%%%%%%%%%%%%%%%%%%%%%

We also studied $r_{H}$ and $1/T$ relations in terms of $M$ 
for above three cases. We first show those 
for $a+8b<0$ in Fig. 7 (a) and (b), respectively. 
We fixed the parameters $a/Q_{e}^{2}=1$, $b/Q_{e}^{2}=-1$, 
$Q_{e}/l_{p}=1$ and $Q_{m}/l_{p}=10^{-4}$, $1$. 
For $Q_{m}/l_{p}=1$, we can easily see specific properties of 
the magnetically charged case though for $Q_{m}/l_{p}=10^{-4}$ 
we can not. But it is not true. Even for $Q_{m}/l_{p}=10^{-4}$, 
there exist solutions in the $r_{H}\to 0$ limit where the temperature 
diverges and $M\to -\infty$. They are clear from Fig. 8 which 
is a magnification of Fig. 7 (a).  

We show corresponding diagrams for $a+8b\geq 0$ 
in Fig. 9 (a) and (b), respectively.
We fixed the parameters $a/Q_{e}^{2}=1$, $b/Q_{e}^{2}=-0.125$ 
(i.e., $a+8b=0$), $-0.1$, $0$, $0.1$, 
$Q_{e}/l_{p}=1$ and $Q_{m}/l_{p}=1$. 
For $b/Q_{e}^{2}=-0.125$, $-0.1$, it is almost indistinguishable 
in this diagram though the electric field has different limit for 
$r_{H}\to 0$ in these two cases. We can see the character 
like the magnetically charged case. But for $b/Q_{e}^{2}=0$, $0.1$, we can 
see the character like the electrically charged case, 
i.e., solutions below an extreme solution do not exist. 
The curve below the points $A$ is a sequence of inner horizons. 
We also investigated those for various $Q_{e}/Q_{m}$ ratio 
which suggest that whether or not solutions 
in the $r_{H}\to 0$ limit exist depends only on $a$, $b$. 
Thus if we believe that this system is realistic, the coupling 
constants decide the final fate of black holes.

%%%%%%%%%%%%%%%%%%%%%%%%
\begin{figure}[htbp]
\segmentfig{9cm}{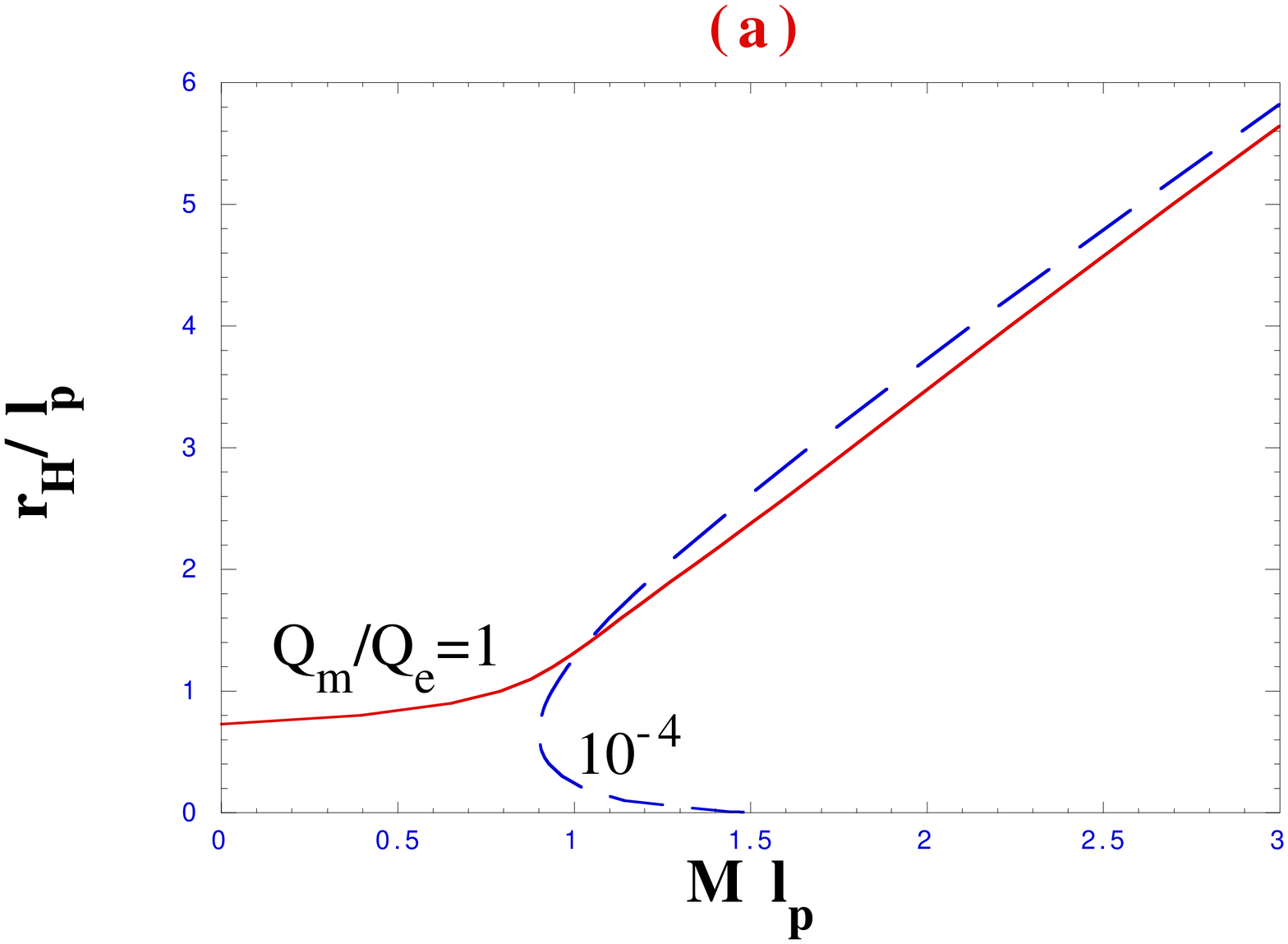}{}  \\
\segmentfig{9cm}{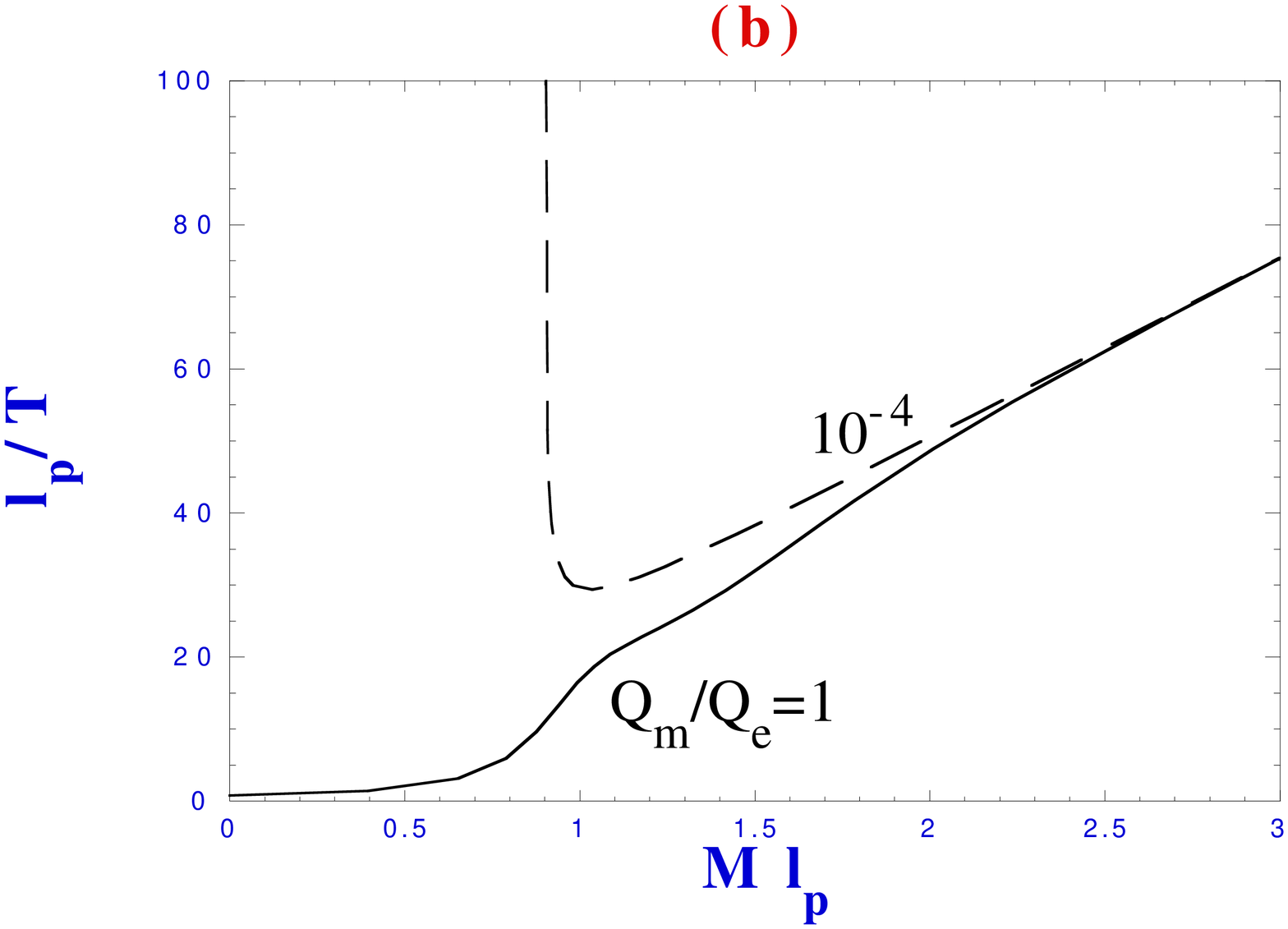}{}
\caption{(a) $M$-$r_H$ (b) $M$-$1/T$ relations for dyon black holes 
for $a/Q_{e}^{2}=1$, $b/Q_{e}^{2}=-1$, (i.e., $a+8b<0$) 
$Q_{e}/l_{p}=1$ and $Q_{m}/l_{p}=10^{-4},1$. It seems that 
solutions in the $r_{H}\to 0$ limit only exist for $Q_{m}/l_{p}=1$. 
But it is not true.  
\label{M-?dyon1}
}
\end{figure}
%%%%%%%%%%%%%%%%%%%%%%%%

%%%%%%%%%%%%%%%%%%%%%%%%
\begin{figure}
\begin{center}
\singlefig{9cm}{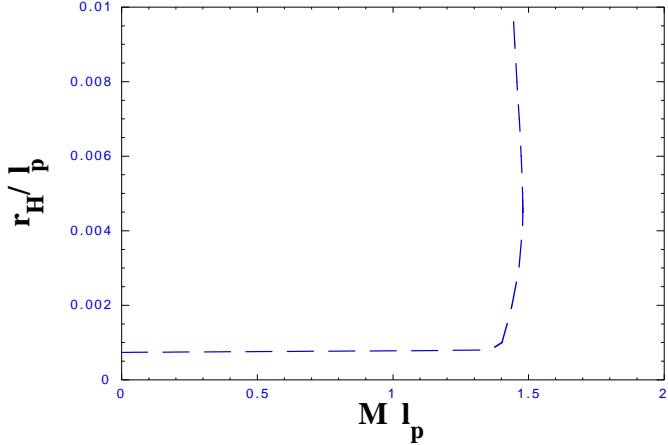}
\caption{Magnification of Fig. 7 (a) which shows that there exist solutions 
in the $r_{H}\to 0$ limit even for $Q_{m}/l_{p}=10^{-4}$.  
\label{M-?dyon2} }
\end{center}
\end{figure}
%%%%%%%%%%%%%%%%%%%%

%%%%%%%%%%%%%%%%%%%%%%%%%%%%%%%%%%%%%%%%%%%%%%%%%%%%%%%%%%%%
%                                                          %
\section{Conclusion and Discussion}    %
%
%%%%%%%%%%%%%%%%%%%%%%%%%%%%%%%%%%%%%%%%%%%%%%%%%%%%%%%%%%%%

We note our conclusions and future work. We investigate black hole 
solutions in the EEH system for electrically charged, 
magnetically charged 
and dyonic solutions. They have remarkable thermodynamical properties. 

(i) For the magnetically charged case, the properties of the black holes 
change qualitatively for $a=a_{crit}$. There is an extreme 
solution only for $a\leq a_{crit}$. There are solutions in 
the $r_H \to 0$ limit for arbitrary $a$ and the temperature 
diverges in this limit.  

(ii) For the electrically charged case which was analyzed previously, 
though the lower limit of the horizon becomes small as we take $a$ 
to be large, the 
final state of the black hole when we consider the evaporating process is 
similar to the RN one. The causality change is also already 
pointed out, i.e., inner horizon appears only for $M<M_0$, as 
we confirmed it. 
But the point $M=M_0$ is not relevant to the change of 
the stability if we 
apply the turning point method. 

(iii) As for the dyon case, we showed that there exists solutions in 
the $r_{H} \to 0$ limit for $a+8b \leq 0$ and approach the magnetially 
charged case. For $a+8b >0$, our results suggest that whether or not 
solutions in the $r_{H} \to 0$ limit exists depends only on $a$ and $b$ 
not on $Q_{m}/Q_{e}$ ratio except for vanishing $Q_{m}$ or $Q_{e}$. 

We now comment on our future work. 
Though we considered the Einstein-Hilbert action as a gravitational part, 
it is important to generalize to think higher order curvature corrections. 
It may be 
interesting to think about black hole solutions in the action which 
generalize the EH action to preserve supersymmetry\cite{who}. 
Our solutions have pathological properties like the negative gravitational 
mass. There may exist mechanism which prevent such properties as 
in \cite{Kallosh}. An other concern we have is to think about 
black hole solutions including 
cosmological term, because its importance is recognized both 
in observatioal and in theoretical perspectives.  

%%%%%%%%%%%%%%%%%%%%%%%
\begin{figure}[htbp]
\segmentfig{9cm}{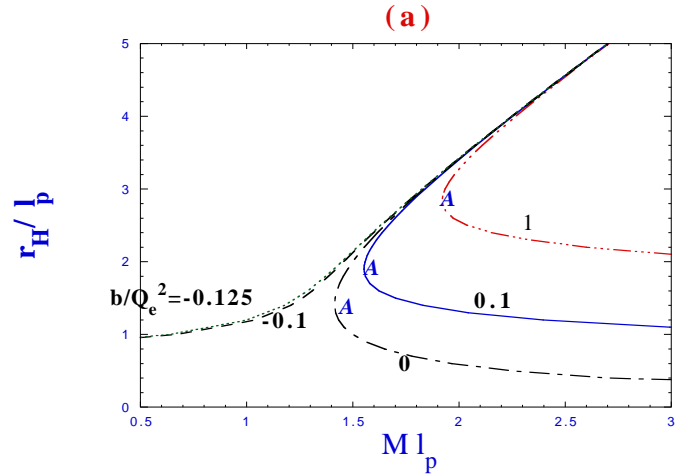}{}  \\
\segmentfig{9cm}{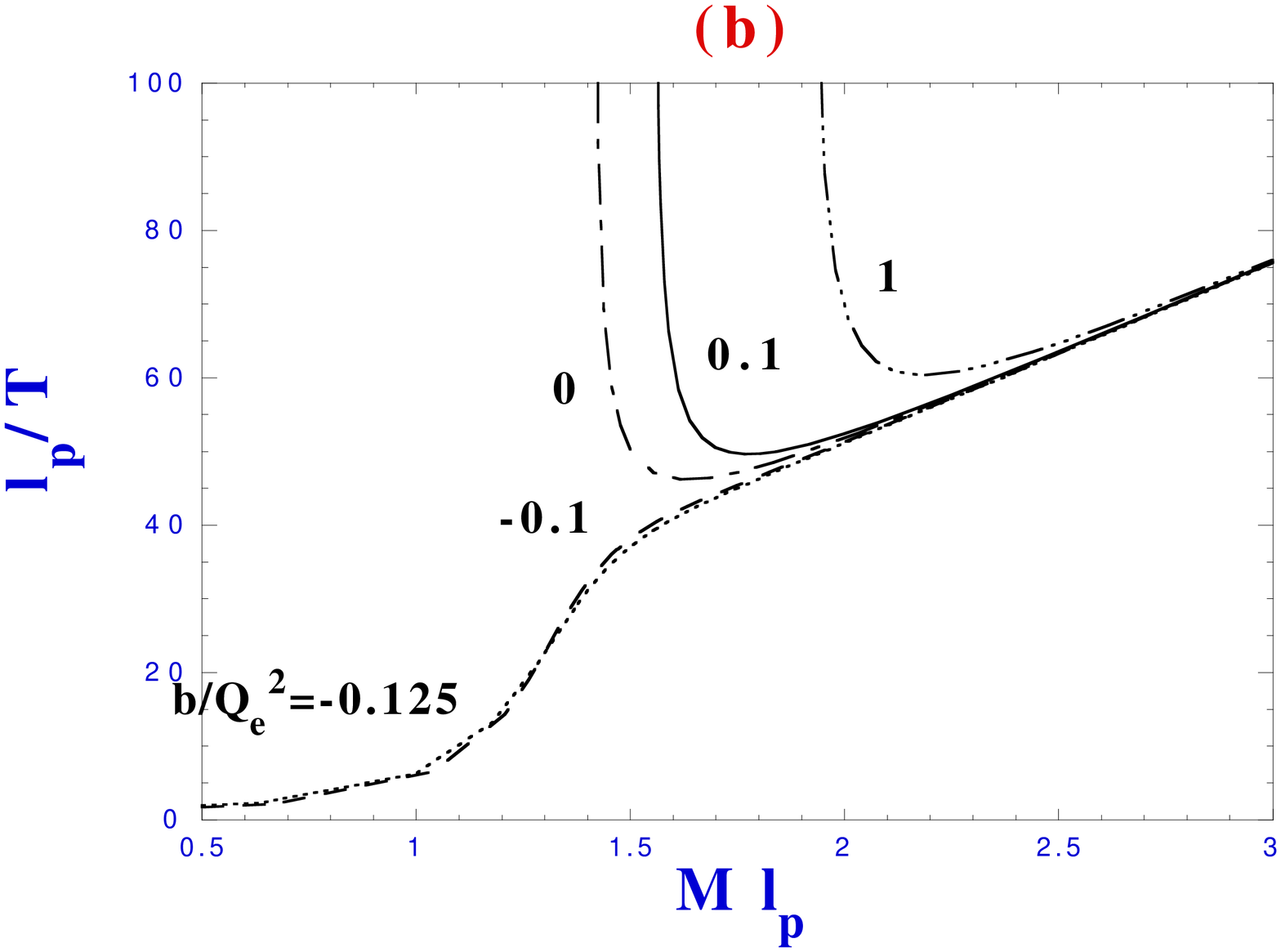}{}
\caption{(a) $M$-$r_H$ (b) $M$-$1/T$ relations for dyon black holes 
for $a/Q_{e}^{2}=1$, $b/Q_{e}^{2}=-0.125$, $-0.1$,  $0$, $0.1$, 
$Q_{e}/l_{p}=1$ and $Q_{m}/l_{p}=1$. 
For $b/Q_{e}^{2}=-0.125$, $-0.1$, these figures are almot indistinguishable 
for these two cases and resemble those 
for the magnetically charged case. 
But for $b/Q_{e}^{2}=0$, $0.1$ 
they resemble those for the electrically charged case. 
\label{M-?dyon3}
}
\end{figure}
%%%%%%%%%%%%%%%%%%%%%%%%
%%%%%%%%%%
\section*{ACKOWLEDGEMENTS}
Special thanks to  G. W. Gibbons, H. Watabe, K. Maeda and T. Torii 
for useful discussions and also thanks to M. James 
for checking our English.  
T. T is thankful for  financial support from the
JSPS. This work was supported partially by a 
JSPS Grant-in-Aid (No. 106613). 

%%%%%%%%%%%%%%%%%%%%%%%%%%%%%%%%%%%%%%%


\begin{thebibliography}{99}
\bibitem{compare}
If we use the subscript $_{Oli}$ to 
represent his variables, $4aA'^{2}=x_{Oli},\ A'(r)=-E_{Oli}(r), 
\ Q_e=-q_{Oli},\ a=-1/(2b_{Oli})$.
\end{thebibliography}
\end{document}